\documentstyle[12pt,aasms4]{article}

\lefthead{Pereira, Braga \& Jablonski}
\righthead{GX1+4}

\begin{document}

\newbox\grsign \setbox\grsign=\hbox{$>$} \newdimen\grdimen \grdimen=\ht\grsign
\newbox\simlessbox \newbox\simgreatbox
\setbox\simgreatbox=\hbox{\raise.5ex\hbox{$>$}\llap
     {\lower.5ex\hbox{$\sim$}}}\ht1=\grdimen\dp1=0pt
\setbox\simlessbox=\hbox{\raise.5ex\hbox{$<$}\llap
     {\lower.5ex\hbox{$\sim$}}}\ht2=\grdimen\dp2=0pt
\def\simgreat{\mathrel{\copy\simgreatbox}}
\def\simless{\mathrel{\copy\simlessbox}}
\newbox\simppropto
\setbox\simppropto=\hbox{\raise.5ex\hbox{$\sim$}\llap
     {\lower.5ex\hbox{$\propto$}}}\ht2=\grdimen\dp2=0pt
\def\simpropto{\mathrel{\copy\simppropto}}
\def\ts{\thinspace}

\title{The Orbital Period of the Accreting Pulsar GX~1+4} 

\author{M.\ G.\ Pereira\altaffilmark{}}
\affil{Divis\~ao de Astrof\'\i sica, Instituto Nacional de Pesquisas
Espaciais, CP~515, 12201--970, S\~ao Jos\'e dos Campos, Brazil\\
Electronic mail: marildo@das.inpe.br}

\author{J.\ Braga\altaffilmark{}}
\affil{Divis\~ao de Astrof\'\i sica, Instituto Nacional de Pesquisas
Espaciais, CP~515, 12201--970, S\~ao Jos\'e dos Campos, Brazil\\
Electronic mail: braga@das.inpe.br}

\and 

\author{F.\ Jablonski\altaffilmark{}}
\affil{Divis\~ao de Astrof\'\i sica, Instituto Nacional de Pesquisas
Espaciais, CP~515, 12201--970, S\~ao Jos\'e dos Campos, Brazil\\
Electronic mail: chico@das.inpe.br}

\begin{abstract}
  
  We report strong evidence for a $\sim$ 304-day periodicity in the
  spin history of the accretion-powered pulsar GX\ts 1+4 that is most
  probably associated with the orbital period of the system. We have
  used data from the Burst and Transient Source Experiment on the {\it
    Compton Gamma Ray Observatory} to show a clear periodic modulation
  of the pulsar frequency from 1991 to date, in excellent agreement
  with the ephemeris proposed by Cutler, Dennis \& Dolan (1986).  Our
  results indicate that the orbital period of GX\ts 1+4 is
  303.8$\pm$1.1\ts days, making it the widest known low-mass X-ray
  binary system by more than one order of magnitude and putting this
  long-standing question to rest. A likely scenario for this system is
  an elliptical orbit in which the neutron star decreases its
  spin-down rate (or even exhibits a momentary spin-up behavior) at
  periastron passages due to the higher torque exerted by the
  accretion disk onto the magnetosphere of the neutron star. These
  results are not inconsistent with both the X-ray pulsed flux light
  curve measured by BATSE during the same epoch and the X-ray flux
  history from the All-Sky Monitor (ASM) onboard the {\it Rossi X-Ray
    Timing Explorer}.

\end{abstract}


\section{Introduction} 

GX~1+4 is a bright Galactic Center accretion-powered pulsar in a
low-mass x-ray binary system (LMXB) discovered in the early 1970s
(Lewin, Ricker \& McClintock 1971). Throughout the 1970s the pulsar
exhibited a spin-up behavior with the pulsation period decreasing from
135\,s to less than 110\,s (Cutler, Dennis \& Dolan 1986 -- hereafter
CDD86 -- and references therein), corresponding to a spin-up rate of
$\dot{P} \sim -2$\,s/year.  After experiencing an extended
low-intensity state in the early 1980s (Hall \& Develaar 1983;
McClintock \& Leventhal 1989), GX~1+4 re-emerged in a spin-down state
(Makishima et al.\ 1988; Sakao et al.\ 1990) with approximately the
same $\mid\dot{P}\mid$ and stayed in this state ever since, with
occasional short-term variations of $\dot{P}$.

Infrared observations and optical spectroscopy of GX~1+4 established a
rare association of a neutron star with a M5~III giant star,
V2116\,Oph, in a symbiotic binary system (Glass \& Feast 1973;
Davidsen, Malina \& Bowyer 1977; Chakrabarty \& Roche 1997). The
identification was made secure by a ROSAT accurate position
determination (Predehl, Friedrich \& Staubert 1995) and by the
discovery of optical pulsations in V2116\,Oph consistent with the spin
period of the neutron star (Jablonski et al.\ 1997, Pereira et al.\
1997). In comparison with the other four known LMXB
accretion-powered pulsars (GRO~J1744$-$28, Her~X-1, 4U~1627$-$67 and
the recently discovered millisecond accreting pulsar SAX J1808.4-3658
-- Wijnands \& van der Klis 1998), GX~1+4 has a much longer (factor of
$\sim$ 100) spin period and its orbital period, albeit not securely
measured until this work, was known to be at least one order of
magnitude longer than the periods of the other systems. Physically
quantitative lower limits on the binary period of GX 1+4 were derived
by Chakrabarty and Roche (1997), who showed that the binary period
must be at least 100~d, and is probably more than 260~d.

In 1991, the Burst and Transient Source Experiment (BATSE) on the
Compton Gamma Ray Observatory (CGRO) initiated a continuous and nearly
uniform monitoring of GX~1+4. The BATSE observations confirmed the
spin-down trend with occasional dramatic spin-up/down torque reversal
events (Chakrabarty 1996, Chakrabarty et al.\ 1997, Nelson et al.\ 1997).
        
Attempts to find the orbital period of GX~1+4 by Doppler shifts of the
pulsar pulse timing or optical lines have both been inconclusive so
far. For the X-ray timing measurements, the accretion torque magnitude
is much larger than the expected orbital Doppler shifts, and the
torque fluctuations have significant power at the time scales
comparable to the expected binary period (Chakrabarty 1996). In the
case of the optical lines, the problem is the long period ($\simgreat$
100 days) expected (Davidsen, Malina \& Bowyer 1977, Doty, Hoffman \&
Lewin 1981; Sood et al.\ 1995).  Long-term optical photometry in R
band has shown variations in the light curve with periods of $\sim 30$
and $\sim 110$\,days (Pereira, Braga \& Jablonski 1996; Pereira
1998). Using a small number of X-ray measurements carried out during
the spin-up phase of GX~1+4 in the 1970s, CDD86 produced an ephemeris
for predicting periodical enhancements in the spin-up rate of the
neutron star. A possible interpretation for this periodic behavior is
that the neutron star and the red giant are in an elliptical orbit
with a 304-day period.
         
In this work we report the results of a comprehensive time-series
analysis of the BATSE data on GX~1+4 in an attempt to find the orbital
period of the system. We discuss the implications that can be
drawn from our results in light of the possible models for this
source and show that the elliptical orbit interpretation is probably
the correct one.  We present a refined version of the ephemeris
originally proposed by CDD86.

\section{Data Analysis and Results} 

The frequency and the pulsed flux data between Julian Day (JD)
2448376.5 and 2451138.5 (i.e., 1991 April 29 to 1998 October 20) used
in this work were obtained from Chakrabarty (1996) and from the BATSE
public domain data available at {\tt
  http://www.batse.msfc.nasa.gov/data/pulsar/sources}. The 20--50~keV
pulsed signals are extracted from DISCLA 1.024s channel 1 data.  After
being weighted according to the aspect angle of each detector to a
source on an existing source list, data are treated to remove the
background variations, barycentered and epoch-folded over a set of
grid points in the vicinity of the expected pulse frequency. Frequency
and modeled flux are inspected daily for significant detections and
the reports are produced twice a week.  15-day mean values for the
fluxes and pulse frequencies of GX~1+4 were calculated for the entire
dataset.

A dataset of GX~1+4 residual pulsation frequencies was obtained from
the frequency history by subtraction of a standard cubic spline
function to remove low frequency variations in the spin-down
trend. The fitting points are mean frequency values calculated over
suitably chosen time intervals. The results of the spline fitting are
fairly insensitive to intervals greater than $\sim$ 200 days between
fitting points (we have used $\Delta t=215$~days). The pulsed X-ray
flux, frequency history and residual frequencies are shown in Fig.~1
as functions of time.

We have carried out a power spectrum analysis to search for
periodicities of less than 1000\,days in both the residual frequency
and the pulsed flux data. A Lomb-Scargle periodogram (Press et al.\
1992), suitable for time series with gaps, shows a significant
periodic signal at 302.0~days (Fig.~2) in the residual frequency time
series. This value is insensitive to any oversampling factor greater
than 2 over the Nyquist interval. The power spectrum shows a red noise
with an approximate power-law index index of $-$2. In order to
estimate the statistical significance of the detection, a series of
numerical simulations of the frequency time series with 1-sigma
gaussian deviations (using the error bars of the data points) were
performed. In the frequency domain of the simulated light curves, we
forced the power spectra to have a power-law index of $-2$, and
transformed the results back to the time domain. We then selected the
times in the simulated light curves to match the observed times for
the datapoints and subtracted the spline with $\Delta t=215$~days. We
finally calculated the Lomb periodogram of the simulated
time-series. The simulations show that the use of the 215-d spline,
besides providing an effective filter for frequencies below $\sim
2\times 10^{-3} {\rm d}^{-1}$, does not produce power in any specific
frequency in the range of interest, i.e., no monocromatic or even QPO
signals are produced. By comparing the amplitude of our 302-day peak
with the local value obtained by the mean of the numerical simulations
(the peak is a factor of 13.91 higher), we obtain a statistical
significance of 99.98\% for the detection. Epoch folding the data
using the 302-day period yields a 1-$\sigma$ uncertainty of 1.7~days.

In the pulsed flux data, the most interesting feature is a wide
structure of low-significance peaks observed in the range
200-500\,days, with no significant peak at $\sim$ 300\ts days.

By analyzing the variation of the period of GX~1+4 during the spin-up
phase in the 1970s, CDD86 proposed a 304\,-day orbital period and an
ephemeris to predict the events of enhanced spin-up: $ T = {\rm JD\ 
  2,444,574.5} \pm 304\; n$, where $n$ is an integer. This ephemeris
is based on four events discussed by the authors, whose existence was
inferred from ad-hoc assumptions and extrapolations of the
observations.  The projected enhanced spin-up events derived from that
ephemeris for the epochs contained in the BATSE dataset, represented
as solid vertical lines in the lower panel of Fig.\,1, are in
excellent agreement with the BATSE reduced spin-down and spin-up
events. The BATSE dataset is obviously significantly more reliable
than CDD86's inasmuch as it is based on 9 well-covered events measured
with the same instrument as opposed to the 4 events discussed in
CDD86. The striking agreement of CDD86's ephemeris with the BATSE
observations is very conspicuous and give a very strong support to the
claim that the orbital period of the system is indeed $\sim$ 304 days.
Taking integer cycle numbers, with the $T0$ epoch of CDD86 as cycle
$-23$, and performing a linear least-squares fit to the frequency
residuals seen in the lower panel of Fig.~1, we find that the
following ephemeris can represent the time of occurrence $T$ of the
maxima in the frequency residuals:

\begin{equation} T = {\rm JD\ 2,448,571.3} (\pm 3.2) \pm 303.8 (\pm
1.1) \; n, \end{equation} 
where $n$ is any integer. The events predicted by the above ephemeris
are shown as vertical dashed lines in the three panels of Fig.\
1. Taking into account a conservative uncertainty estimate of 30 days
for the peaks of the BATSE events, the reduced $\chi^2$ of the fit is
$\chi^2_{\hbox{r}} = 0.61$. The value of $303.8\pm 1.1$~days for the
orbital period is consistent with the one obtained through power
spectrum analysis performed on the BATSE data, which gives further
support for the period determination.

\section{Discussion}

In the BATSE era, the long term frequency history of GX\ts 1+4 shown
in Fig.~1 (middle panel) exhibits a characteristic spin-down trend
with an average rate of $\sim 1.8$\,s/year. Frequency derivative
reversals occur on times preceding the epochs of events labeled \# 5,
7 and 9 in the bottom panel. The upper panel of Fig.~1 shows that
these events are somewhat correlated with rather intense flares in the
pulsed flux. In the 1970s, when the measurements used by CDD86 were
carried out, the source was in a spin-up extended state. The scenario
proposed by CDD86 to explain the periodic occurrence of enhanced
spin-up events was that the system was in a elliptical orbit and the
periastron passages would occur when $\dot{P}$ is maximum, as expected
in standard accretion from a spherically expanding stellar wind. It is
widely accepted today, as inferred from GX~1+4's optical/IR properties
(Jablonski et al.\ 1997; Chakrabarty, van Kerkwijk \& Larkin 1998;
Chakrabarty et al.\ 1997; Chakrabarty \& Roche 1997), that the system
has an accretion disk.

Since the neutron star is currently spinning-down, the radius at which
the magnetosphere boundary would corotate with the disk, $r_{\rm co} =
(G M P^2/4\pi^2)^{1/3} \sim 3.6 \times 10^4 \;\; P_{\rm 100s}^{2/3}$
km, where $M$ is the mass of the neutron star (assumed to be $\approx$
1.4~M$_{\odot}$) and $P_{\rm 100s}$ is the spin period in units of
100\ts seconds, is probably smaller than the magnetosphere radius $r_M
\sim 4.1 \times 10^4 \;\; L_{36}^{-2/7}$~ km, where $L_{36}$ is the
X-ray luminosity in units of 10$^{36}$ erg/s (Frank, King \& Raine
1992). This value for $r_M$ assumes a surface magnetic field of $\sim$
10$^{14}$~G for GX~1+4 (Makishima et al.\ 1988, White 1988,
Chakrabarty et al.\ 1997, Cui 1997). Since the pulse period is $\sim$
120~s and the luminosity is typically $\simless 10^{37}$~erg/s, the
period is close to the equilibrium value, for which $r_{\rm co} \sim
r_M$. This allows spin-down to occur even though accretion continues,
the centrifugal barrier not being sufficiently effective (White 1988).

Assuming that the elliptical orbit is the correct interpretation for
the origin of the modulation, the mass accretion rate (and hence the
luminosity) should increase as the neutron star approaches periastron,
making $r_M$ approach $r_{\rm co}$. As the velocity gradient between
the disk material and the material flowing along the magnetic field
lines decreases, the spin-down torque gets smaller and the neutron
star decelerates at a slower rate. We expect that this mechanism will
produce a peak in the frequency residuals close to the periastron
epoch, beyond which the neutron star will start to get back to a
higher spin-down rate. Occasionally, due to the highly variable mass
loss rate of the red giant, $r_{\rm co}$ will surpass $r_M$ and the
neutron star will {\it spin-up\/} for a brief period of time during
periastron, as observed in the BATSE frequency curve in events 5, 7
and 9. According to this picture, one would expect an increase in
X-ray luminosity at periastron.  Although this is only marginally
indicated in the BATSE pulsed flux light curve, it should be pointed
out that total flux data from the All Sky Monitor (ASM) onboard RXTE
for the epoch MJD 50088 to 51044 does not correlate significantly with
the BATSE pulsed flux, indicating that the pulsed flux may not be a
good tracer of the accretion luminosity in this system. Furthermore,
the periodic $\sim 5\mu$Hz excursions in the residual frequency would
lead to very low-significance variations in the X-ray flux measured by
the ASM, as we now show. Taking the fiducial torque $N_{0} = \dot{M}
\sqrt{G M_{\rm X} r_{\rm co}}$ given by Bildsten et al.\ (1997), where
$\dot{M}$ is the accretion rate and $M_{\rm X}$ is the mass of the
neutron star, as an order-of-magnitude estimation (since $r_{\rm co}
\approx r_{\rm M}$), we can establish a lower limit to the variation
in $\dot{M}$ ($\Delta\dot{M}$) that produced the residual torque,
using the fact that $\dot{\nu} = N_{0}/2\pi I$, where $I$ is the
moment of inertia of the neutron star (Ravenhall \& Pethick 1994).
Since the relative variation in flux ($F$) scales as the relative
variation in luminosity, we get $\Delta F/F \sim 0.3$ for $L \sim
10^{37}$ erg/s. The typical ASM GX\ts 1+4 flux is $\sim 1\pm 2$
count/s in the 2--10\ts keV, so the expected variations of $\sim$
0.3\ts counts/s would be very hard to detect, given the available
observational coverage.  This is consistent with the lack of any
significant periodic signal in our calculation of the power spectrum
of the entire available ASM flux history of GX 1+4 (from MJD 50088 to
51353).  In the elliptical orbit interpretation, one would also expect
that tidal torques exerted by the red giant envelope would circularize
the orbit in a short time scale (Verbunt \& Phinney 1995). However,
with a period of $\sim$ 300 days, the red giant radius is probably
less than 7\% of the binary separation, as shown below. Since the rate
scales as $(R_{\rm c}/a)^{-8}$, where $R_{\rm c}$ is the red giant
radius and $a$ is the binary separation, we do not expect the
circularization time scale to be too short. Furthermore, Verbunt \&
Phinney (1995) show that for orbital periods longer than about 200
days, the eccentricities of red giant binaries in open clusters span
the full range.

An alternative interpretation for the observed modulation would be the
presence of oscillation modes in the red giant star. For an M5 giant,
persistent radial oscillations with a period of $\sim$ 300\ts days are
perfectly plausible (Whitelock 1987). In this case, the oscillations
could excite a modulation in the mass loss rate through the stellar
wind that could produce the modulated torque history. However, the
stability of the infrared magnitudes of V2116\ts Oph (Chakrabarty \&
Roche 1997) preclude it from being a long-period variable, since these
stars undergo regular $\simgreat$~1~mag variations in the infrared
(Whitelock 1987). In addition, the secular optical light curve in the
$R$ band obtained by our group at Laborat\'orio Nacional de Astrof\'\i
sica (Brazil) from 1991 to date shows no signs of these oscillations
(Pereira, Braga \& Jablonski 1996; Pereira 1998).

It is noteworthy that the amplitude of the residual frequency
oscillations in GX\ts 1+4 cannot be attributed to Doppler shifts
(Chakrabarty 1996). A firm lower limit for the companion mass is given
by the X-ray mass function $f_{\rm X} (M) = (c \Delta\nu/\nu)^3 P_{\rm
  orb}/2 \pi G$, which would be equal to $\sim$ 210~M$_{\odot}$ for a
$\sim 5 \mu$Hz amplitude and a 304-day orbital period. This is clearly
too massive for a red giant and actually for any stellar companion.
There is also no evidence of Doppler shifts in the spectral lines of
V2116 Oph (Sood et al.\ 1995; Chakrabarty \& Roche 1997), which could
be an indication that the inclination of the system is fairly low.

The spectral and luminosity classification of V2116\ts Oph, together
with the measured interstellar extinction of $A_V \approx 5$, is
consistent with a low-mass star (M $\sim$ 0.8$-$2\ts M$_{\odot}$) on
the first-ascent red giant branch at a distance of 3$-$6~kpc
(Chakrabarty \& Roche 1997). The range of radii for such a start is
$\sim 50-110$\ts R${\odot}$. The size of the Roche lobe of this
object as the companion in the binary system can be estimated by the
radius of a sphere with the same volume as the lobe,

\begin{equation} R_L = 1.42 \times 10^{11} \; M_{\rm X}^{1/3} \;
\frac{(1+q)^{1/3} q^{2/3}}{0.6 q^{2/3} + \ln (1+q^{1/3})} \;\; P_d
\;\; {\rm cm}, 
\end{equation} 
where $q = M_g/M_{\rm X}$ is the mass ratio of the red giant and the
neutron star, $M_{\rm X}$ is in solar mass units and $P_d$ is the
orbital period in days (Eggleton 1983). Assuming $M_{\rm
X} = 1.4$ and $P_{\rm d} = 304$, the range of values obtained for
$R_L$ is 546\ts R${\odot}-$780\ts R${\odot}$ for GX\ts 1+4, with the
binary separation ranging from 1640 to 1890\ts R$\odot$. Thus, the
companion is probably not filling its Roche lobe and the accretion
disk forms from the slow, dense stellar wind of the red giant.

Another interesting argument leading to a $\sim$ 300-day orbital
period for GX\ts 1+4 comes from the work of van Paradijs \& McClintock
(1994), according to which the absolute visual magnitudes of low-mass
X-ray binary systems seem to correlate linearly with the quantity
$\Sigma = P_{\rm orb}^{2/3}\gamma^{1/2}$, where $\gamma = L_{\rm
X}/L_{\rm Edd}$ is the accretion luminosity in units of the Eddington
luminosity. Taking the value of $M_V \approx -4.2$ for the disk light
not contaminated by H$\alpha$ obtained by Jablonski \& Pereira (1997),
we get $P_{\rm orb} \sim 270\pm 82$\ts days for $L_{\rm X} \sim L_{\rm
Edd}$, which is fully consistent with our results. It should be noted, 
however, that this model is based upon the assumption that the optical
emission is dominated by reprocessing of X-rays in the accretion disk, 
which is not clear to be the case in GX\ts 1+4.

In conclusion, we have shown that the long-sought orbital period of
GX\ts 1+4 is very likely to be 304~days, as proposed in 1986 by CDD86
with marginal confidence. A more thorough covering of the X-ray
luminosity of the system, with high sensitivity and spanning several
cycles, will be very important to test the elliptical orbit model.

\vspace{0.3cm}

We thank Dr.\ Bob Wilson from NASA Marshall Space Flight Center for
gently providing us BATSE frequency and flux data on GX\ts 1+4.  M.\
P.\ is supported by a FAPESP Postdoctoral fellowship at INPE under
grant 98/16529-9. J.\ B.\ thanks CNPq for support under grant
300689/92-6. F.\ J.\ acknowledges support by PRONEX/FINEP under grant
41.96.0908.00. We thank an anonymous referee for very important
corrections and suggestions. 
 

\clearpage

\figcaption{{\it Upper panel\/}: Light curve of the 20-50\ts keV
pulsed flux of GX\ts 1+4 as measured by BATSE from 1991 to 1998; {\it
middle panel\/}: GX\ts 1+4 frequency measurements by BATSE over the
same period. The error bars are in general smaller than the size of
the dots. The solid curve is a cubic spline fit to the data; {\it
lower panel\/}: frequency residuals. The dotted vertical lines mark
the times times predicted by the ephemeris calculated in this work,
whereas the solid vertical lines show the predictions according to
CDD86's ephemeris. The events of positive residual frequency
modulation are labeled for reference in the text.}

\figcaption{Lomb-Scargle periodogram of the frequency residuals of
GX\ts 1+4 from 1991 to 1998, represented by the histogram-type solid
line. The standard solid line is the mean of 1500 numerical
simulations carried out in order to calculate the significance level
of the detection. The upper dotted line indicates a significance level
of 0.001, whereas the lower dotted line indicates a significance level
of 0.01. }

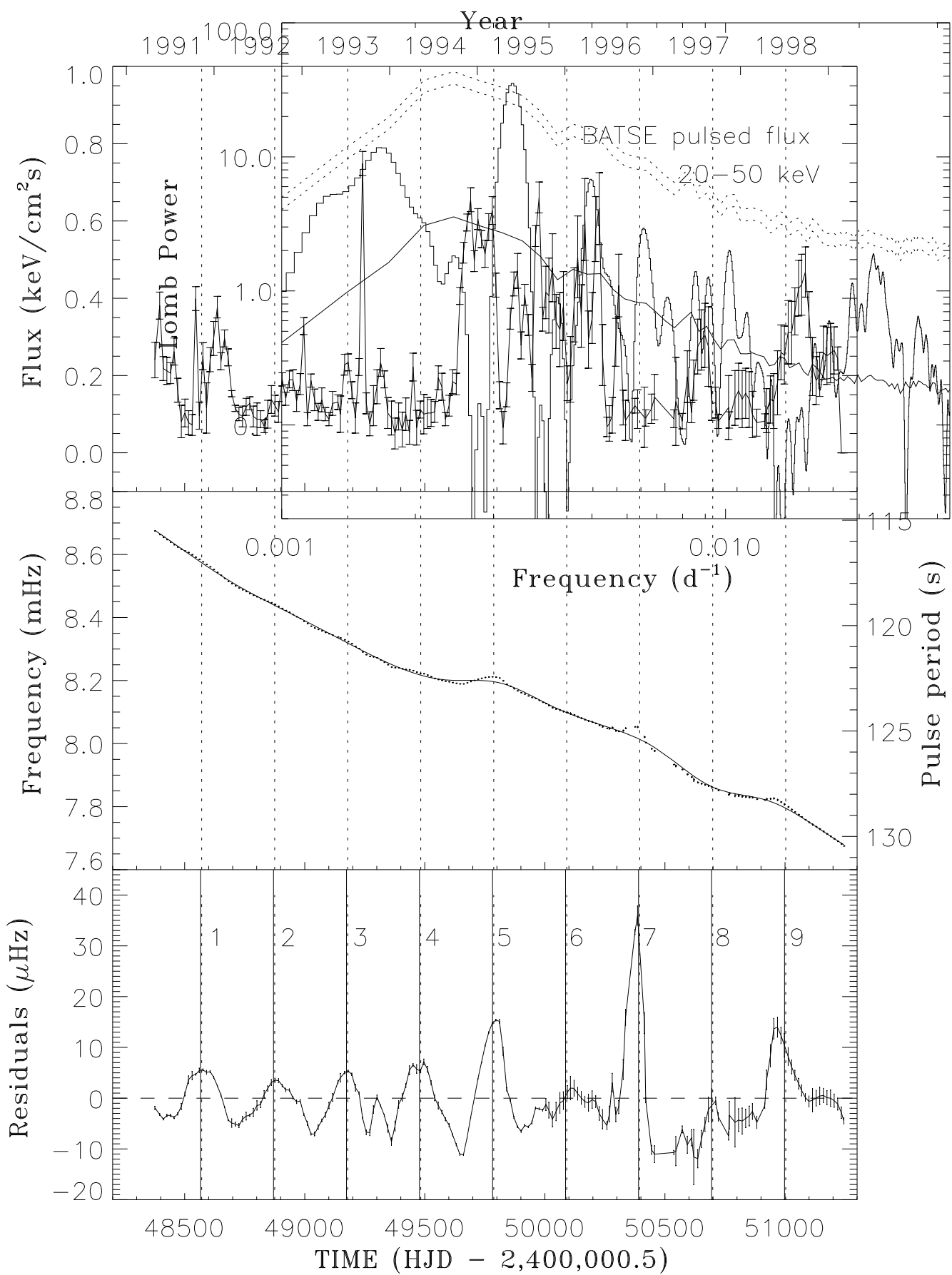
\end{document}